\documentstyle[12pt,prd,aps]{revtex}

\begin{document}

\def\abstract#1{\begin{center}{\bf ABSTRACT}\end{center}
\par #1}
\def\title#1{\begin{center}{\large {#1}}\end{center}}
\def\author#1{\begin{center}{\sc #1}\end{center}}
\def\address#1{\begin{center}{\it #1}\end{center}}

\def\pubnum{98--8}

\begin{titlepage}
\hfill
\parbox{6cm}{{YITP--\pubnum} \par Jan. 1998}
\parbox{6cm}{}
\par
\vspace{1.5cm}
\begin{center}
\Large
New varieties of Gowdy spacetimes
\end{center}
\vskip 1cm
\author{Masayuki TANIMOTO\footnote{JSPS Research Fellow. 
Electronic mail: tanimoto@yukawa.kyoto-u.ac.jp}}

\address{Yukawa Institute for Theoretical Physics,
        Kyoto University, Kyoto 606-8502, Japan.}
\vskip 1 cm

\abstract{ Gowdy spacetimes are generalized to admit two commuting
  spatial {\it local} Killing vectors, and some new varieties of them
  are presented, which are all closely related to Thurston's geometries.
  Explicit spatial compactifications, as well as the boundary conditions
  for the metrics are given in a systematic way. A short comment on an
  implication to their dynamics toward the initial singularity is made.}


\end{titlepage} \addtocounter{page}{1}

\def\R{{\bf R}}
\def\Z{{\bf Z}}
\def\x{{\bf x}}
\def\del{\partial}
\def\Lap{\bigtriangleup}
\def\Div{{\rm div}\ }
\def\rot{{\rm rot}\ }
\def\curl{{\rm curl}\ }
\def\grad{{\rm grad}\ }
\def\Tr{{\rm Tr}}
\def\^{\wedge}
\def\goinf{\rightarrow\infty}
\def\goes{\rightarrow}
\def\bm{\boldmath}
\def\-{{-1}}
\def\inv{^{-1}}
\def\sqr{^{1/2}}
\def\isqr{^{-1/2}}

\def\reff#1{(\ref{#1})}
\def\vb#1{{\partial \over \partial #1}} 
\def\Del#1#2{{\partial #1 \over \partial #2}}
\def\Dell#1#2{{\partial^2 #1 \over \partial {#2}^2}}
\def\Dif#1#2{{d #1 \over d #2}}
\def\Lie#1{ {\cal L}_{#1} }
\def\diag#1{{\rm diag}(#1)}
\def\abs#1{\left | #1 \right |}
\def\rcp#1{{1\over #1}}
\def\paren#1{\left( #1 \right)}
\def\brace#1{\left\{ #1 \right\}}
\def\bra#1{\left[ #1 \right]}
\def\angl#1{\left\langle #1 \right\rangle}
\def\vector#1#2#3{\paren{\begin{array}{c} #1 \\ #2 \\ #3 \end{array}}}
\def\svector#1#2{\paren{\begin{array}{c} #1 \\ #2 \end{array}}}
\def\matrix#1#2#3#4#5#6#7#8#9{
        \left( \begin{array}{ccc}
                        #1 & #2 & #3 \\ #4 & #5 & #6 \\ #7 & #8 & #9
        \end{array}     \right) }

\def\GL#1{{\rm GL}(#1)}
\def\SL#1{{\rm SL}(#1)}
\def\PSL#1{{\rm PSL}(#1)}
\def\O#1{{\rm O}(#1)}
\def\SO#1{{\rm SO}(#1)}
\def\IO#1{{\rm IO}(#1)}
\def\ISO#1{{\rm ISO}(#1)}
\def\U#1{{\rm U}(#1)}
\def\SU#1{{\rm SU}(#1)}

\def\diffeos{diffeomorphisms}
\def\diffeo{diffeomorphism}
\def\Teich{{Teichm\"{u}ller }}
\def\Poin{{Poincar\'{e} }}

\def\Gam{\mbox{$\Gamma$}}
\def\d{{\rm d}}
\def\VII#1{\mbox{VII${}_{#1}$} }
\def\VI#1{\mbox{VI${}_{#1}$} }
\def\Nil{{\rm Nil}}
\def\Sol{{\rm Sol}}

\def\hh{{h}}
\def\gg{{\rm g}}
\def\uh#1#2{\hh^{#1#2}}
\def\dh#1#2{\hh_{#1#2}}
\def\ug#1#2{\gg^{#1#2}}
\def\dg#1#2{\gg_{#1#2}}
\def\uug#1#2{\tilde{\gg}^{#1#2}}
\def\udg#1#2{\tilde{\gg}_{#1#2}}
\def\udh#1#2{\tilde{\hh}_{#1#2}}

\def\MM{\four{M}}
\def\uMM{\four{\tilde{M}}}
\def\Mtil{\tilde M}

\def\s#1{\sigma^{#1}}
\def\a#1#2{a_{#1}{}^{#2}}
\def\p#1#2{p^{#1}{}_{#2}}
\def\dug#1#2{g_{#1}{}^{#2}}

\def\wa{&=&}
\def\wb{&\equiv&}

\section{Introduction}

Spatially compact inhomogeneous spacetimes admitting two commuting
spatial Killing vectors are known as Gowdy spacetimes \cite{Gow}, which
are recently paid large attention as favorable models for the studies of
the asymptotic behavior toward the cosmological initial singularity
\cite{IM,GM,BM,HS,BG}. Since Gowdy spacetimes provide the simplest
inhomogeneous cosmologies, it seems natural to use them in such a kind
of studies first developed based on Bianchi homogeneous cosmologies
\cite{LL} by Belinski et.al. \cite{BKL}. This article's aim is not,
however, to elucidate the dynamics of Gowdy spacetimes, but to point out
some new features pertaining to their varieties, which may open new
windows to look at the initial singularity.

As is well known, Bianchi homogeneous cosmologies are classified by the
local structures of simply transitive three dimensional groups, Bianchi
I$\sim$IX groups. Spatially compact Bianchi cosmologies
\cite{KTH,TKH1,TKH2} are more diverse than the open ones, since possible
compact topologies are in general very diverse for each Bianchi
(universal covering) cosmology. Nevertheless, Gowdy spacetimes having
lower symmetry than these homogeneous ones are known to have very
limited varieties, i.e., only $T^3$, $S^2\times S^1$, $S^3$, and lens
spaces as their possible spatial topologies. Why is their diversity so
poor?  In fact, all the Bianchi cosmologies except types VIII and IX
admit two commuting Killing vectors, and moreover, compactifications are
possible except for types IV and VI${}_{a\neq0}$ \cite{Fa,KTH}. Since
these compact Bianchi models can be thought of as, if exist, homogeneous
limits of compact inhomogeneous models which admit two commuting Killing
vectors, it seems that Gowdy models should be more diverse. The solution
to this paradox is in whether the Killing vectors are local or global.
That is, the restriction for the possible topologies of Gowdy spacetimes
is a consequence of the definition that the two commuting Killing
vectors must be {\it globally} defined, while spatially compact Bianchi
cosmologies admit in general only {\it local} Killing vectors.

If we consider Killing vectors for the simplification of Einstein's
equation, which is local in nature, the imposition of the globality of
the Killing vectors is evidently unnecessary. We therefore generalize
Gowdy spacetimes in this article to admit two commuting {\it local}
Killing vectors. (This type of generalization was also considered by
Rendall \cite{Ren} in a different approach from ours.)  We will actually
find that there exist rich (topologically infinitely many) varieties of
Gowdy spacetimes.

As is well known in theory of three dimensional topology, Thurston
\cite{Th} enumerated eight types of homogeneous 3-manifolds, called {\it
  model geometries}, $S^3$, $E^3$, $H^3$, $S^2\times E^1$, $H^2\times
E^1$, $\widetilde{SL}(2,{\bf R})$, nilgeometry(Nil), and
solvegeometry(Sol), and proved in essence that any compact three
dimensional manifold which admit a locally homogeneous Riemannian metric
is a compact quotient of one of these eight types of homogeneous
manifolds.  (See Sec.3.8 of Ref.\cite{Th}.)  We effectively utilize
these model geometries, as in the compact locally homogeneous cases
\cite{KTH,TKH1}.

In the next section, we show how we can apply Thurston's theorem to find
(generalized) Gowdy spacetimes, and in the subsequent two sections
present three new types of Gowdy spacetimes as examples. The final
section is devoted to conclusions, including a comment on the
dynamics of the three Gowdy spacetimes.

\section{Possible topologies}

To find possible topologies of Gowdy spacetimes, we consider a
`homogenization' of a compact Riemannian 3-manifold
{\mbox{$(M,h_{ab})$}}\ which admit two commuting local Killing vectors.
Namely, suppose we can smoothly deform the metric $h_{ab}$ preserving
the two local Killing vectors and can take a locally homogeneous limit
$(M,h_{ab}^{\rm lh})$ of the Gowdy space {\mbox{$(M,h_{ab})$}}.  Since
the universal cover {\mbox{$(\tilde M,\tilde{h}_{ab}^{\rm lh})$}}\ of
{\mbox{$(M,h_{ab}^{\rm lh})$}}\ is homogeneous, it must be one of the
BKSN types, i.e., Bianchi homogeneous 3-manifolds and the
Kantowski-Sachs-Nariai (KSN) homogeneous 3-manifold \cite{Na,KS}.  On
the other hand, since the homogenization {\mbox{$(M,h_{ab}^{\rm lh})$}}
is compact, {\mbox{$(M,h_{ab}^{\rm lh})$}} must be homeomorphic to a
compact quotient of one the eight Thurston's model geometries.

A (model) geometry is the pair $({\tilde M},G)$ of a manifold ${\tilde
  M}$ and a group $G$ of diffeomorphisms\ on ${\tilde M}$, such that $G$
acts transitively on ${\tilde M}$ with compact isotropy subgroups.
Since $G$ acts transitively on ${\tilde M}$, $({\tilde M},G)$ can be
thought of as an equivalence class of the homogeneous manifolds whose
isometry group is isomorphic to (or includes a subgroup isomorphic to)
$G$ \cite{TKH1}.  We for convenience call $({\tilde M},G)$ a {\it
  subgeometry} of $({\tilde M},G')$ if $G$ is a subgroup of other
transitive group $G'$ with compact isotropy subgroups.  Moreover, if
geometry $({\tilde M},G)$ is not a subgeometry of any geometry, then we
call $({\tilde M},G)$ a {\it maximal} geometry, and if $({\tilde M},G)$
does not have any subgeometry, then we call $({\tilde M},G)$ a {\it
  minimal} geometry.  While Thurston's eight geometries are maximal
geometries, all the BKSN types except Bianchi IV and VI${}_{a\neq0}$
can be thought of as the minimal geometries of Thurston's
geometries \cite{KTH,Fa}. This correspondence helps apply Thurston's
geometries to spacetime models.

The group $G$ of the geometry $({\tilde M},G)$ which corresponds to
{\mbox{$(\tilde M,\tilde{h}_{ab}^{\rm lh})$}} should include a subgroup
isomorphic to one of ${\bf R}^2$, ${\bf R}\times U(1)$, and $U(1)\times
U(1)$, corresponding to the commuting two Killing vectors.  All
Thurston's geometries satisfy this condition.  More precisely, what
include ${\bf R}^2$ are geometries $E^3$, $H^3$, $H^2\times E^1$,
$\widetilde{SL}(2,{\bf R})$, Nil, and Sol.  Geometry $S^2\times E^1$
includes ${\bf R}\times U(1)$, and $S^3$ does $U(1)\times U(1)$.
Moreover, all the minimal geometries of these Thurston's geometries
except Bianchi VIII ($\widetilde{SL}(2,{\bf R})$ as the maximal one) and
IX ($S^3$, similarly) also satisfy this condition.  Any compact quotient
for all the homogeneous 3-manifolds except Bianchi VIII and IX can
therefore be a homogenization of a Gowdy space.  Even for Bianchi VIII
and IX, they can be a homogenization if imposing a fourth Killing vector
on them.

This observation tells what homogeneous manifold can be the universal
cover of a homogeneous limit of a Gowdy space.  Once fixed such a
homogenization with isometry group $G$, the universal cover of the
corresponding Gowdy space can be determined as follows.  That is, we may
find the subgroup $H$ of $G$ such that the actions are smooth along the
two commuting Killing vectors, but discrete along the third one.  The
$H$-invariant metric {\it is} the universal cover metric of the Gowdy
space (or spacetime), if the fundamental group can be represented in
$H$.  In fact, the $H$-invariance is a necessary condition for the
universal cover to admit a (spatially) compact quotient, so that we need
to check that.  As we will see, it is easy to write such an
$H$-invariant metric if using the invariant 1-forms for the
corresponding Bianchi type. The KSN type has already been discussed in
Ref.\cite{Gow}, so we do not consider it in this article.

In the following, we show how these ideas work by presenting three
examples, which are all new.

\section{Gowdy on Nil$\times{\bf R}$}

As a first example let us consider Nil, which possesses Bianchi II as
its minimal geometry. Note that the Bianchi II homogeneous spaces are
characterized by the following commutation relations for the three
Killing vectors
\begin{equation}
        [\xi_1,\xi_2]=-\xi_3,\; [\xi_2,\xi_3]=0,\; [\xi_3,\xi_1]=0.
        \label{II-2}
\end{equation}
In terms of coordinate basis they can be represented by
\begin{equation}
        \xi_1={\partial \over \partial x}+y{\partial \over \partial z},\;
 \xi_2={\partial \over \partial y},\; \xi_3={\partial \over \partial z}.
        \label{II-1}
\end{equation}
For future use, we take this opportunity to write the finite actions
generated by these $\xi_i$'s
\begin{equation}
        \left(\begin{array}{c} a \\ b \\ c
          \end{array}\right)\left(\begin{array}{c} x \\ y \\ z
          \end{array}\right)=
\left(\begin{array}{c} {a+x} \\ {b+y} \\ {c+z+ay} \end{array}\right),
        \label{II-6}
\end{equation}
where the component vectors
\begin{equation}
        \left\{ \left(\begin{array}{c} a \\ b \\ c \end{array}\right)
        \equiv e^{c\xi_3}e^{b\xi_2}e^{a\xi_1} \bigg|
        \; a,b,c\in{\bf R}\right\}
        \label{II-6a}
\end{equation}
form Bianchi II group $G_{{\rm II}}$.  ($e^{a\xi_i}$ denotes the
one-parameter group of diffeomorphisms generated by $\xi_i$.)  Note that
the ``spatial point'' $(x,y,z)$ is the image of the origin $(0,0,0)$ by
the ``group element'' $(x,y,z)$, so that Eq.\reff{II-6} itself gives the
multiplication rule in Bianchi II group.

From the commutation relations (\ref{II-2}), we find two possible
choices of obtaining a Gowdy spacetime model, i.e., one (Type 1) is to
keep $\xi_2$ and $\xi_3$ as the two commuting Killing vectors and
consider inhomogeneity along $\xi_1$, and the other (Type 2) is to keep
$\xi_1$ and $\xi_3$ as Killing vectors and consider inhomogeneity along
$\xi_2$.  We consider the Type 1 first.

Note that the invariant 1-forms of Bianchi II, given by
\begin{equation}
        \sigma^1={\rm d}x,\; \sigma^2={\rm d}y,\; \sigma^3={\rm d}z-x {\rm d}y
        \label{II-3}
\end{equation}
are globally defined on $\tilde M={\bf R}^3$, so that we can expand any
metric on $\tilde M$ by these 1-forms
\begin{equation}
        {\rm d}l^2=h_{ij} \sigma^ i \sigma^ j,
        \label{II-5}
\end{equation}
with globally defined metric functions $h_{ij}$.  Note also that the
subgroup $H_{{\rm II}}$ of $G_{{\rm II}}$ of which action is discrete
along $\xi_1$ is formed by
\begin{equation}
        \left\{
          \left(\begin{array}{c} {2m\pi} \\ b \\ c \end{array}\right)
          \bigg|\; m\in{\bf Z},\; b,c\in{\bf R} \right\}.
        \label{II-8}
\end{equation}
(The choice of the interval for the first component is arbitrary. Our
choice of $2\pi$ is just for definiteness.) Since the invariant 1-forms
(\ref{II-3}) are invariant under $H_{{\rm II}}$ (and $G_{{\rm II}}$ by
definition), the metric (\ref{II-5}) is invariant under $H_{{\rm II}}$
iff so are the metric functions $h_{ij}$.  This requirement is
equivalent to that $h_{ij}$'s depend only on $x$ and are periodic with
period $2\pi$, i.e.,
\begin{equation}
         h_{ij}=h_{ij}(x)=h_{ij}(x+2\pi).
        \label{II-dh}
\end{equation}
The homogeneous limit can be achieved when $h_{ij}={\rm constants}$.
So, we have found the inhomogeneous metric of (the universal cover of) a
Gowdy space, given by Eq.(\ref{II-5}) with the boundary condition
(\ref{II-dh}).

Now, we can write down the ``appropriate'' spacetime metric by
\cite{Gow} imposing the two-surface orthogonality and choosing the
isothermal coordinates for the reference surface.  After all, we obtain
the spacetime metric
\begin{equation}
        {\rm d}s^2 =
        e^{-\lambda/2}t^{-1/2}(-{\rm d}t^2+(\sigma^1)^2)+
        R\big[e^P(\sigma^2)^2+
        2e^PQ\sigma^2\sigma^3+(e^PQ^2+e^{-P})(\sigma^3)^2\big],
        \label{II-9}
\end{equation}
where $P,Q,R$ and $\lambda$ are functions of $t$ and $x$ and are
periodic in $x$ with period $2\pi$. We have followed the parameterization
of a recent paper \cite{BG} to make comparisons easier.  The isometry
group for the spacetime is unchanged, given by $H_{{\rm II}}$ (\ref{II-8}).

This metric should be considered as the universal cover metric
$\tilde{{\rm g}}_{ab}$ of a Gowdy spacetime {\mbox{$(M\times{\bf R},{\rm
 g}_{ab})$}}.  Any topology of $M$ is possible if the fundamental
group $\pi_1(M)$ can be represented in $H_{\rm II}$.  One can choose
such a $\pi_1(M)$ in the list of compact quotients modeled on Nil,
presented in Ref.\cite{KTH}.  For example, choose the manifold
``$b/1(n)$'', characterized by
\begin{equation}
        \pi_1(M_n)=\big\langle g_1,g_2,g_3; [g_1,g_2]=g_3^n,
        [g_1,g_3]=1, [g_2,g_3]=1 \big\rangle,
        \label{II-10}
\end{equation}
where $n$ is a positive integer parameterizing the family $b/1$ of
topologies on Nil.  Then using the multiplication rule \reff{II-6}, we
can find the representation up to conjugations
\begin{eqnarray}
        \Gamma_n &=& \left\{ g_1,g_2,g_3 \right\} \nonumber \\
        &=& \left\{
          \left(
            \begin{array}{c} {2p\pi} \\ {g_{1}{}^{2}} \\ {0} \end{array}
          \right),
          \left(
         \begin{array}{c} {2q\pi}\\{g_{2}{}^{2}}\\{g_{2}{}^{3}}
         \end{array}
       \right),
       \left(\begin{array}{c}
           {0}\\{0}\\{{2\pi\over n}(pg_{2}{}^{2}-qg_{1}{}^{2})} 
         \end{array}\right)\right\},
        \label{II-11}
\end{eqnarray}
where $p,q\in{\bf Z}$, and $g_{1}{}^{2},g_{2}{}^{2},g_{2}{}^{3}\in{\bf
  R}$.  Thus, we have obtained the Gowdy spacetime $(M_n\times{\bf
  R},{\rm g}_{ab})=$ $(\tilde M_n\times{\bf R},\tilde{{\rm
    g}}_{ab})/\Gamma_n$, where the action of {\mbox{$\Gamma_n$}}\ is
defined by Eq.(\ref{II-6}).  We may think of the real parameters
$(g_{1}{}^{2},g_{2}{}^{2},g_{2}{}^{3})$ as the moduli parameters for the
spacetime.

The Type 2 can be obtained by interchanging the roles of $\s1$ and
$\s2$. For example, the metric can be obtained from Eq.\reff{II-9} by
transforming $\s1\goes\s2$, $\s2\goes\s1$, $P(t,x)\goes P(t,y)$,
etc.. The isometry group thereof is formed by the actions
\begin{equation}
        \left\{
          \left(\begin{array}{c} a \\ {2m\pi} \\ c \end{array}\right)
          \bigg|\; m\in{\bf Z},\; a,c\in{\bf R} \right\}.
        \label{II-8b}
\end{equation}
As can be easily checked, we can represent the fundamental group
\reff{II-10} into this isometry group in a form similar to
Eq.\reff{II-11}.

Let us consider the vacuum Einstein equations. To be specific, we
consider the Type 1. (As for the final results, we will present for both
Types 1 and 2.) Note that if neglecting the boundary conditions, our
metric (\ref{II-9}) is essentially the same as the metric given in
Ref.\cite{Gow}
\begin{equation}
  \label{eq:goworg}
  {\rm d}s^2 = 
        e^{-\lambda/2}t^{-1/2}(-{\rm d}t^2+{\rm d}x^2)+R\big[e^P{\rm d}y^2+
        2e^PQ{\rm d}y{\rm d}z+(e^PQ^2+e^{-P}){\rm d}z^2\big],
\end{equation}
where $P$, $Q$, $R$, and $\lambda$ are functions of $t$ and $x$.
In fact, our metric \reff{II-9} is obtained from this one by transforming
\begin{equation}
  \label{eq:gowtrans}
        P \rightarrow P+\ln [(1-xQ)^2+x^2e^{-2P}], \quad
        Q \rightarrow \frac{Q(1-xQ)^2-xe^{-2P}}{(1-xQ)^2+x^2e^{-2P}}
\end{equation}
(, though this does not preserve the periodicity of $P$ and $Q$).
Moreover, the role of $R$ as the area function of the group orbits,
consisting of flat $T^2$'s \cite{note2}, is the same for both metrics.
This can be checked by noticing that the natural volume element of the
second term in the metric \reff{II-9} is given by $R\,\d y\^\d z$.  As a
result, the function $R$ of our metric satisfies the same key equation
as that of metric \reff{eq:goworg}, i.e., \cite{Gow}
\begin{equation}
        \partial_{tt} R- \partial_{xx} R=0.
        \label{II-12}
\end{equation}
This equation can also be checked by a direct substitution into the
vacuum Einstein equation. Then, since the group orbits for our spatial
manifold do not degenerate everywhere (except at the initial
singularity), $R$ can be taken as $t(\equiv e^{-\tau})$, as in the Gowdy
model on $T^3\times\R$ \cite{Gow}. $t=0$ $(\tau=+\infty)$ corresponds to
the initial singularity.  With this choice of $R$, the remaining
independent Einstein's equations for our metric \reff{II-9} are found by
a direct calculation to be
\begin{eqnarray}
        \ddot P - e^{-2\tau}P''-e^{2P}\dot Q^2+
        e^{-2\tau}\big[ e^{2P}(Q^2\pm Q')^2\pm 2Q'-e^{-2P} \big] \wa 0,
        \nonumber \\
        \ddot Q - e^{-2\tau}Q''+2\dot P\dot Q-
        2e^{-2\tau}\big[ P'Q'\pm (P'\mp Q)(Q^2+e^{-2P}) \big] \wa 0, 
        \label{II-13}
\end{eqnarray}
and
\begin{eqnarray}
        \lambda' &-&
        2(P'\mp 2Q)\dot P\mp 2\big[ e^{2P}(Q^2\pm Q')-1 \big]\dot Q
        = 0,
        \nonumber \\
        \dot\lambda &-&
        \dot P^2 -e^{2P}\dot Q^2- e^{-2\tau}\big[ e^{2P}(Q^2\pm Q')^2
        +P'^2+2Q^2 \mp2(Q'+2P'Q)+e^{-2P} \big] = 0,
        \label{II-14}
\end{eqnarray}
where dot and dash denote $\tau$ and $x$(or $y$ for the Type 2)
derivatives, respectively. The upper and lower signs are for the Type 1
and Type 2, respectively. Note that $P$ and $Q$ are not constrained,
since $\lambda$ does not appear in Eqs.(\ref{II-13}). This is one of the
advantages of Gowdy models.  The integrability condition for the
constraint equations (\ref{II-14}) for $\lambda$ is automatically
satisfied with Eqs.(\ref{II-13}). The boundary conditions for $P$, $Q$,
and $\lambda$ are the spatially periodic ones. The Hamiltonian for the
dynamical equations \reff{II-13} can be guessed from straightforwardly
reducing the Einstein-Hilbert action with the metric \reff{II-9}. It is
given by
\begin{equation}
  \label{eq:II-ham}
  H=\rcp2\int_0^{2\pi}\!\d\theta\, [\pi_P^2+e^{-2P}\pi_Q^2]+ e^{-2\tau}
  \bra{(P'\mp 2Q)^2-2(Q^2\pm Q')+e^{2P}(Q^2\pm Q')^2+e^{-2P}},
\end{equation}
where $\pi_P$ and $\pi_Q$ are the conjugate momenta of $P$ and $Q$,
respectively. The integration measure $\theta$ is $x$ for the Type 1, or
$y$ for the Type 2.

\section{Gowdy on Sol$\times{\bf R}$ and VII${}_0\times{\bf R}$}

Next examples correspond to Bianchi VI${}_0$ and VII${}_0$ as minimal
geometries. Their maximal geometries are Sol and $E^3$, respectively.
Since the local structures for Bianchi VI${}_0$ and VII${}_0$ are
similar, we treat them in parallel in this section. The basic procedure
is the same as that of the previous section.

We first observe the commutation relations of the sets of the three
Killing vectors of the Bianchi groups
\begin{eqnarray}
        [\xi_1,\xi_2]\wa 0,\; [\xi_2,\xi_3]=-\xi_1,\; [\xi_3,\xi_1]=\xi_2\;
        \mbox{: VI${}_0$},  \label{VI-2} \\
        {[\xi_1,\xi_2]}\wa 0,\; [\xi_2,\xi_3]=-\xi_1,\; [\xi_3,\xi_1]=-\xi_2\;
        \mbox{: VII${}_0$}.  \label{VII-2}
\end{eqnarray}
(We write ``tags'' to distinguish the two types as above.)  In terms of
coordinate basis,
\begin{eqnarray}
        \xi_1 \wa
        {\partial \over \partial x}+{\partial \over \partial y},\;
        \xi_2={\partial \over \partial x}-{\partial \over \partial y},\;
        \xi_3 = 
        {\partial \over \partial z}-x{\partial \over \partial x}+
        y{\partial \over \partial y}\; \mbox{: VI${}_0$},
        \label{VI-1} \\
        \xi_1 \wa
        {\partial \over \partial x},\;
        \xi_2={\partial \over \partial y},\;
        \xi_3 = 
        {\partial \over \partial z}-y{\partial \over \partial x}+
        x{\partial \over \partial y}\; \mbox{: VII${}_0$}.
        \label{VII-1}
\end{eqnarray}
The Bianchi groups are formed by the finite actions of $\xi_i$'s, which
we take as
\begin{eqnarray}
  G_{{\rm VI}_0}\wa\left\{\left(\begin{array}{c} a \\ b \\ c
      \end{array}\right)\equiv
    e^{{a\over2}(\xi_1+\xi_2)}e^{{b\over2}(\xi_1-\xi_2)}e^{c\xi_3}\bigg|
    \; a,b,c\in{\bf R}\right\},
        \label{VI-6a} \\
  G_{{\rm VII}_0}\wa\left\{\left(\begin{array}{c} a \\ b \\ c
      \end{array}\right)\equiv
    e^{a\xi_1}e^{b\xi_2}e^{c\xi_3}\bigg|
    \; a,b,c\in{\bf R}\right\}.
        \label{VII-6a}
\end{eqnarray}
The actions or multiplications are given by
\begin{equation}
        \left(\begin{array}{c}
            a \\ b \\ c 
          \end{array}\right)
        \left(\begin{array}{c}
            x \\ y \\ z \end{array}\right) =
        \left(\begin{array}{c}
            {a+e^{-c}x}\\{b+e^cy}\\{c+z}
        \end{array}\right)\mbox{: VI${}_0$},\;
        \left(\begin{array}{c}
            a \\ b \\ c 
          \end{array}\right)
        \left(\begin{array}{c}
            x \\ y \\ z \end{array}\right) =
        \svector{\svector{a}{b}+{\cal R}_c\svector{x}{y}}{c+z}
        \mbox{: VII${}_0$},
        \label{VI-6}
\end{equation}
where ${\cal R}_c$ is the rotation matrix by angle $c$.

From the commutation relations (\ref{VI-2}) and \reff{VII-2}, we find
sole possibility of obtaining the Gowdy spaces, the ones inhomogeneous
along $\xi_3$.  Using the invariant 1-forms
\begin{eqnarray}
        \sigma^1 \wa {1\over\sqrt2}\big( e^z{\rm d}x+e^{-z}{\rm d}y \big),\;
        \sigma^2={1\over\sqrt2}\big( -e^z{\rm d}x+e^{-z}{\rm d}y \big),\;
        \sigma^3 = {\rm d}z\, \mbox{: VI${}_0$},
        \label{VI-3} \\
        \sigma^1 \wa  \cos z{\rm d}x+\sin z{\rm d}y,\;
        \sigma^2= -\sin z{\rm d}x+\cos z{\rm d}y,\;
        \sigma^3 = {\rm d}z\, \mbox{: VII${}_0$},
        \label{VII-3}
\end{eqnarray}
and the periodic functions
\begin{eqnarray}
        P \wa P(\tau,z)=P(\tau,z+z_0),\; 
        Q=Q(\tau,z)=Q(\tau,z+z_0), \nonumber \\
        \lambda \wa \lambda(\tau,z)=\lambda(\tau,z+z_0),
        \label{VI-PQl}
\end{eqnarray}
we can immediately write the same spacetime metric for the two types as
\begin{equation}
        {\rm d}s^2 = 
        e^{-\lambda/2}e^{\tau/2}(-e^{-2\tau}{\rm d}\tau^2+(\sigma^3)^2)+
        e^{-\tau}\big[e^P(\sigma^1)^2+
        2e^PQ\sigma^1\sigma^2+(e^PQ^2+e^{-P})(\sigma^2)^2\big].
        \label{VI-4}
\end{equation}
As we will see later, the period $z_0$ is specified depending upon the
topology. The isometry groups $H_{{\rm VI}_0}$ and $H_{{\rm VII}_0}$ for
the spacetime metric (\ref{VI-4}) is formed by the discrete actions
along $\xi_3$ and can be written in the same form
\begin{equation}
        \left\{
\left(\begin{array}{c} {a}\\{b}\\{m z_0} \end{array}\right) \bigg|\;
          a,b\in{\bf R},\; m\in{\bf Z} \right\}.
        \label{VI-5}
\end{equation}

In the following, we consider the vacuum Einstein equations first, since 
the compactifications must be discussed separately for each universal
cover.

In the metric (\ref{VI-4}), we wrote the area function of the group
orbit as $e^{-\tau}=t$, since, as in the Nil case, the group orbits do
not degenerate except at the initial singularity for both types. The
remaining Einstein equations are then found to be
\begin{eqnarray}
        \ddot P &-&
        e^{-2\tau}P''-e^{2P}\dot Q^2+ e^{-2\tau}\big[ 
        e^{2P}(Q'+Q^2\mp1)^2+2Q'-e^{-2P} \big]=0,
        \nonumber \\
        \ddot Q  &-&
        e^{-2\tau}Q''+2\dot P\dot Q-
        2e^{-2\tau}\big[ P'Q'+(P'-Q)(e^{-2P}+Q^2\mp1) \big]=0,
        \label{VI-ein}
\end{eqnarray}
and
\begin{eqnarray}
        \lambda' &-& 2(P'-2Q)\dot P-2\big[ e^{2P}(Q'+Q^2\mp1)-1 \big]
        \dot Q=0, \nonumber \\
        \dot\lambda &-& \dot P^2-e^{2P}\dot Q^2- 
        e^{-2\tau}\big[(P'-2Q)^2-2(Q'+Q^2\mp1)+e^{2P}(Q'+Q^2\mp1)^2
        +e^{-2P} \big]=0,
        \label{VI-con}
\end{eqnarray}
where dot and dash denote, respectively, $\tau$ and $z$ derivatives.
The upper and lower signs are for Sol(VI${}_0$) and VII${}_0$, respectively.
The integrability condition for the constraint equations (\ref{VI-con})
for $\lambda$ is automatically satisfied with Eqs.(\ref{VI-ein}).
The Hamiltonian for the dynamical equations \reff{VI-ein} is given by
\begin{equation}
  \label{eq:vi-ham}
  H=\rcp2\int_0^{z_0}\!\d z\, [\pi_P^2+e^{-2P}\pi_Q^2]+ e^{-2\tau}
  \bra{(P'-2Q)^2-2(Q'+Q^2)+e^{2P}(Q'+Q^2\mp1)^2+e^{-2P}}.
\end{equation}

{\it Compactification for Gowdy on Sol$\times\R$}: We can take any
topology of $M$ of the Gowdy spacetime $(M\times{\bf R},{\rm g}_{ab})$
on Sol$\times\R$ if we can represent the fundamental group $\pi_1(M)$
into $H_{{\rm VI}_0}$ (\ref{VI-5}) together with, if needed, the
disconnected components defined with the discrete isometry
\begin{equation}
        h: (x,y,z)\rightarrow (-x,-y,z).
        \label{VI-h}
\end{equation}
As an explicit example, we consider the major sequence of compact
quotients presented in Ref.\cite{KTH}.  The fundamental groups are
parameterized by an integer $n$ such that $|n|>2$, and given by
\begin{equation}
        \pi_1(M_n) =
        \big\langle g_1,g_2,g_3; [g_1,g_2]=1,\;g_3g_1g_3^{-1}=g_2,\;
        g_3g_2g_3^{-1}=g_1^{-1}g_2^n \big\rangle.
        \label{VI-7}
\end{equation}
(This does not, however, exhaust all the compact quotients \cite{Kod}.)
We find that the representations are, up to conjugations, given by
\begin{equation}
  \label{VI-rep1}
  {\mbox{$\Gamma$}}_{n}=\left\{
\left(\begin{array}{c} {\alpha u_1}\\{\beta u_2}\\{0} \end{array}\right),
\left(\begin{array}{c} {\alpha v_1}\\{\beta v_2}\\{0} \end{array}\right),
\left(\begin{array}{c} 0\\0\\{z_0} \end{array}\right),
\right\}
\end{equation}
for $n>2$, and
\begin{equation}
  \label{VI-rep2}
  {\mbox{$\Gamma$}}_{n}=\left\{
\left(\begin{array}{c} {\alpha u_1}\\{\beta u_2}\\{0} \end{array}\right),
\left(\begin{array}{c} {\alpha v_1}\\{\beta v_2}\\{0} \end{array}\right),
h\circ\left(\begin{array}{c} 0\\0\\{z_0} \end{array}\right),
\right\}
\end{equation}
for $n<-2$, with $\alpha,\beta\in{\bf R}$.  In these representations,
$(u_1,v_1)$, $(u_2,v_2)$, and $z_0$ are determined in such a way that
${\rm sign}(n)e^{-z_0}$ and ${\rm sign}(n)e^{z_0}$ are the eigenvalues
of matrix $\left( \begin{array}{cc} 0 & 1 \\ -1 & n \end{array}
\right)$, and the corresponding normalized eigenvectors are $(u_1,v_1)$
and $(u_2,v_2)$, respectively.  In particular, $
e^{z_0}=|n+\sqrt{n^2-4}|/2 $.  We thus have obtained the Gowdy spacetime
$(M_n\times{\bf R},{\rm g}_{ab})=(\tilde M_n\times{\bf R},\tilde{{\rm
    g}}_{ab})/{\mbox{$\Gamma$}}_n$, where the universal cover metric
$\tilde{{\rm g}}_{ab}$ is given by Eq.(\ref{VI-4}) with 1-forms
\reff{VI-3}.

{\it Compactification for Gowdy on VII${}_0\times\R$}: The Gowdy model
which is most frequently picked up in the literature is the one on
$T^3\times\R$, which would have been obtained by our procedure starting
from Bianchi I. However, another $T^3\times\R$ model can be obtained
from Bianchi VII${}_0$, which we pick up here.

The fundamental group of $T^3$ is the infinite group with three
commuting generators;
\begin{equation}
  \label{eq:pi1t3}
  \pi_1(T^3)=\angl{g_1,g_2,g_3;[g_1,g_2]=1,\; [g_2,g_3]=1,\; [g_3,g_1]=1}.
\end{equation}
The general solution of the representation into ``$G_{{\rm VII}_0}$'' has
already given in Eq.(97) of Ref.\cite{KTH};
\begin{equation}
  \label{eq:gamt3}
  \Gam=\brace{\vector{\dug 11}{\dug12}{2l\pi},
    \vector{\dug 21}{\dug22}{2m\pi},\vector{\dug 31}{\dug32}{2n\pi}},
\end{equation}
where $\dug11$, $\dug12$, $\dug21$, $\dug22$, $\dug31$, $\dug32 \in\R$,
and $l,m,n\in\Z$.  We immediately notice that this representation is
effective even in ``$H_{{\rm VII}_0}$'' if we choose the period $z_0$ to
$2\pi$. So, this representation with $z_0=2\pi$ is the general solution.
The only difference from the locally homogeneous case is that there are
no effective conjugations. The representation \reff{eq:gamt3} therefore
gives the final form of the covering group. Finally, we have obtained
the Gowdy spacetime $(T^3\times{\bf R},{\rm g}_{ab})=(\R^4,\tilde{{\rm
    g}}_{ab})/\Gamma$, where the universal cover metric $\tilde{{\rm
    g}}_{ab}$ is given by Eq.(\ref{VI-4}) with 1-forms \reff{VII-3}.

\section{Conclusions}

We have generalized Gowdy spacetimes to admit two commuting {\it local}
Killing vectors. By this generalization, we gained rich varieties of new
Gowdy spacetimes, but any advantages the original has, e.g., the
simplicity of the vacuum Einstein equations, have not been lost. In this
sense, the original definition demanding global existence of the two
commuting Killing vectors was too much restricted. Our generalization is
natural and useful for physical applications.

We have presented three new Gowdy spacetimes, on Nil$\times{\bf R}$, on
Sol$\times{\bf R}$, and on VII${}_0\times{\bf R}$, which are closely
related to Thurston's geometries, Nil, Sol, and $E^3$, respectively. We
have given not only the vacuum Einstein equations with boundary
conditions but also an explicit representation of the covering group for
each case. These three new models and the one called $T^3\times\R$ model
in the literature have common features like (1) the group orbits do not
degenerate everywhere except at the initial singularity, and (2) there
are two dynamical variables (i.e., $P$ and $Q$). These features make the
four models very similar. In fact, the only essential difference is the
boundary conditions for the metric.  Since we wrote the spacetime
metrics in a suitable way for each case, their Einstein equations look
different each other, but the boundary conditions for the metric
functions are the same, simply periodic.  We could have wrote the
spacetime metric in a common form, e.g., like Eq.\reff{eq:goworg}, but
in that case, the boundary conditions for the metric functions would
have taken inconvenient forms, as we have seen in the case of Nil.

It is worth noting that we can interpret the difference of the boundary
conditions for the metrics as the difference of the ``background
metrics''. For example, the spatial metric of the (conventional)
$T^3\times\R$ model can be smoothly deformed locally flat, so that we
can think that the background is flat, while, say, the spatial metric of
the Nil$\times\R$ model cannot be deformed locally flat but locally
Bianchi II, so that the background is the Bianchi II locally homogeneous
curved space in this case.

Here, we comment on the dynamics near the initial singularity of our
three models.  The Gowdy spacetime on $T^3\times{\bf R}$ is conjectured
\cite{GM} to be asymptotically velocity term dominated (AVTD) \cite{IM}.
Recent investigation by Berger and Garfinkle \cite{BG} supports this,
{\it except for} measure-zero nongeneric spatial points.  They succeeded
to explain the phenomenon by a potential picture, and showed the
nongeneric points correspond to the points where $Q'=0$.  In our new
models, such points correspond to points such that $Q'\pm Q^2=0$ (for
the Types 1 and 2 of Nil) and $Q'+Q^2\mp1=0$ (for Sol and
VII${}_0$). Note that the points where $Q'=0$ is inevitable, since $Q$
is periodic, but our four conditions are not necessarily satisfied in
any spatial point.  One may therefore expect that the Gowdy
Nil$\times{\bf R}$, Sol$\times{\bf R}$, and VII${}_0\times\R$ models are
AVTD {\it everywhere}, so the ``curved backgrounds'' improve the AVTD
behavior.

As a final remark, an extension of our method to the so-called $U(1)$
models \cite{Mon} and other similar ones would also be possible, which
is now under development, as well as a complete classification and
further study of Gowdy models.

\bigskip

{\it Note after the completion of this work}: Recently, Weaver,
Isenberg, and Berger \cite{WIB} applied a Gowdy model on Sol$\times\R$
with magnetic field to examine the Mixmaster behavior toward the initial
singularity. The boundary conditions imposed there is consistent with
ours, though they look different, since they used a metric similar to
Eq.\reff{eq:goworg}.

\section*{Acknowledgments}
The author thanks Prof.~Rendall for drawing his attention to
Ref.\cite{Ren}.  He also acknowledges financial support from the Japan
Society for the Promotion of Science and the Ministry of Education,
Science and Culture.



\begin{thebibliography}{99}
\bibitem{note2}
The group actions are well defined on $(\tilde M,\tilde h_{ab})$, and
the orbits form Euclid planes $E^2$. The compactification of $(\tilde
M,\tilde h_{ab})$ compactifies these $E^2$ to two-tori $T^2$, which we
call the group orbits as well, though the group actions are defined only
locally on the quotient.
\end{thebibliography}
\end{document}